\documentclass[%
 reprint,
unsortedaddress,
 amsmath,amssymb,
 aps,
]{revtex4-2}

\usepackage[english]{babel}  % 或使用 polyglossia 包
\usepackage{epsfig} 
\usepackage{graphicx}% Include figure files
\usepackage{dcolumn}% Align table columns on decimal point
\usepackage{bm}% bold math
\usepackage[colorlinks=true]{hyperref}
\begin{document}

\preprint{APS/123-QED}

\title{Dimensionality Mismatch Enables Decoupled Heat and Charge Transport}% Force line breaks with \\

\author{Luman Shang}
%\email{wuyu@njnu.edu.cn}
\affiliation{Advanced Thermal Management Technology and Functional Materials Laboratory, Ministry of Education Key Laboratory of NSLSCS, School of Energy and Mechanical Engineering, Nanjing Normal University, Nanjing 210023, P. R. China}

\author{Shuming Zeng}
\affiliation{College of Physics Science and Technology, Yangzhou University, Jiangsu, Yangzhou 225009, China}

\author{Chenhan Liu}
\email{chenhanliu@njnu.edu.cn}
\affiliation{Advanced Thermal Management Technology and Functional Materials Laboratory, Ministry of Education Key Laboratory of NSLSCS, School of Energy and Mechanical Engineering, Nanjing Normal University, Nanjing 210023, P. R. China}

\author{Yu Wu}
\email{wuyu@njnu.edu.cn}
\affiliation{Advanced Thermal Management Technology and Functional Materials Laboratory, Ministry of Education Key Laboratory of NSLSCS, School of Energy and Mechanical Engineering, Nanjing Normal University, Nanjing 210023, P. R. China}

%\author{Yu Wu}
%%\email{wuyu@njnu.edu.cn}
%\affiliation{Advanced Thermal Management Technology and Functional Materials Laboratory, Ministry of Education Key Laboratory of NSLSCS, School of Energy and Mechanical Engineering, Nanjing Normal University, Nanjing 210023, P. R. China}
%
%\author{Shuming Zeng}
%\affiliation{College of Physics Science and Technology, Yangzhou University, Jiangsu, Yangzhou 225009, China}
%
%\author{Chenhan Liu}
%\email{chenhanliu@njnu.edu.cn}
%\affiliation{Advanced Thermal Management Technology and Functional Materials Laboratory, Ministry of Education Key Laboratory of NSLSCS, School of Energy and Mechanical Engineering, Nanjing Normal University, Nanjing 210023, P. R. China}
%
%\author{Hao Zhang}
%\affiliation{College of Future Information Technology and Department of Optical Science and Engineering and State Key Laboratory of Photovoltaic Science and Technology, Fudan University, Shanghai 200433, China}
%
%\author{Liujiang Zhou}
%%\email{ljzhou@uestc.edu.cn}
%\affiliation{School of Physics, State Key Laboratory of Electronic Thin Films and Integrated Devices, University of Electronic Science and Technology, Sichuan, Chengdu 610054, China}
%
%\author{Ying Chen}
%\affiliation{Eastern Institute of Technology, Zhejiang, Ningbo 315200, China}
%
%\author{Su-Huai Wei}
%\email{suhuaiwei@eitech.edu.cn}
%\affiliation{Eastern Institute of Technology, Zhejiang, Ningbo 315200, China}

\begin{abstract}
Decoupling heat and charge transport is a key challenge in thermoelectrics. Here, we identify a route to spatially separate phonon and carrier transport in quasi-one-dimensional materials through high-throughput screening of the Materials Project database. Representative Sn$_2$S$_3$ and SbTeI exhibit a strong-intrachain--weak-interchain bonding hierarchy that favors phonon propagation along the chains while suppressing transverse lattice heat transport. In contrast, transverse valence-band states provide effective interchain electronic coupling and relatively light hole transport. This mismatch between lattice and electronic transport dimensionalities produces an inverted thermal--electrical anisotropy. Across the screened candidates, interchain lattice thermal conductivity is strongly suppressed, whereas hole transport remains weakly anisotropic or even favors the interchain direction. For SbTeI, this decoupling yields a maximum $zT$ of approximately 2.1 near 900~K. These results establish dimensionality mismatch as a general strategy for decoupling phonon and carrier transport in thermoelectric materials.
\end{abstract}

\maketitle
\newcommand{\Rmnum}[1]{\uppercase\expandafter{\romannumeral#1}}
\section*{\Rmnum{1}.INTRODUCTION}

Thermoelectric materials enable the direct interconversion between heat and electricity and are therefore attractive for waste-heat recovery, solid-state cooling, and sustainable energy technologies.\cite{Burnete2022,Qin2022,Zheng2022,Crane2004} Their energy-conversion efficiency is commonly characterized by the dimensionless figure of merit $zT=S^2\sigma T/(\kappa_e+\kappa_L)$, where $S$, $\sigma$, $\kappa_e$, and $\kappa_L$ denote the Seebeck coefficient, electrical conductivity, electronic thermal conductivity, and lattice thermal conductivity, respectively. Achieving a high $zT$ requires efficient carrier transport together with suppressed lattice heat conduction. These requirements, however, are often difficult to satisfy simultaneously because electronic and phononic transport are both governed by the underlying crystal structure, chemical bonding, and lattice response. Structural modifications that strongly scatter phonons may also impede carrier motion, whereas strongly connected bonding networks favorable for carrier transport can simultaneously facilitate lattice heat conduction. Decoupling charge and phonon transport therefore remains a central challenge in thermoelectric materials design.

A wide variety of strategies have been developed to suppress $\kappa_L$, including defect engineering\cite{Mao2018,Liu2018a,Poudel2008}, alloying, nanostructuring\cite{Hsu2004,Biswas2012,Hochbaum2008,Singh2024}, interface engineering, and hierarchical structural design. Although these approaches can efficiently enhance phonon scattering and shorten phonon mean free paths, the additional structural disorder may simultaneously degrade carrier transport. This has motivated increasing interest in materials with intrinsically low lattice thermal conductivity, in which heat transport is suppressed by inherent structural and bonding characteristics rather than by extrinsic disorder. Weak chemical bonding, structural complexity\cite{Liu2022,JiaJun2019,Wu2025}, lone-pair electrons\cite{Nielsen2013,Skoug2011,Maria2023}, rattling\cite{Tadano2015,Lin2016,Mukhopadhyay2018} or localized vibrations\cite{Duan2016,Ji2023,Tang2024}, antibonding states\cite{Yuan2023,Yuan2022}, and strong lattice anharmonicity have all been identified as microscopic origins of intrinsically low $\kappa_L$\cite{Wu2024a}. The key challenge, however, is not simply to minimize lattice thermal conductivity, but to suppress phonon transport while preserving favorable electronic conduction.

Low-dimensional crystals provide an additional spatial degree of freedom for addressing this challenge. Layered thermoelectrics have demonstrated that weakly bonded directions can strongly suppress lattice heat transport while still supporting efficient carrier conduction when the electronic structure and carrier-scattering mechanisms are appropriately optimized.\cite{Su2022} These studies establish structural anisotropy as an effective platform for phonon--electron decoupling. However, they primarily seek to maintain or enhance carrier transport along a direction that is already favorable for suppressing heat conduction. A more fundamental possibility remains much less explored: whether phonons and charge carriers can intrinsically prefer different crystallographic directions within the same pristine crystal.

Quasi-one-dimensional (Q-1D) materials provide a natural setting for realizing such directional separation. Their structures are characterized by strongly bonded atomic chains connected through much weaker interchain interactions. This strong-intrachain--weak-interchain bonding hierarchy imposes pronounced mechanical anisotropy on the lattice: the stiff intrachain network facilitates phonon propagation along the chain direction, whereas weak transverse coupling suppresses interchain heat transport. Electronic transport, however, is not determined solely by the mechanical connectivity of the lattice. Instead, it depends on the symmetry, spatial orientation, and hybridization of electronic states near the band edges. Frontier orbitals extending transverse to the chains can therefore establish substantial electronic coupling between neighboring chains even when the corresponding mechanical coupling is weak. Structural one-dimensionality thus does not necessarily imply electronic one-dimensionality.

This distinction opens a transport regime beyond the conventional picture of simply combining low thermal conductivity and high electrical conductivity along the same crystallographic direction. In a Q-1D material, the preferred directions of phonon and carrier transport may instead become inverted: phonons propagate more efficiently along the strongly bonded chains, whereas holes can be transported more efficiently across them through transversely extended electronic states. We refer to this behavior as inverted thermal--electrical transport anisotropy. Microscopically, it originates from a mismatch between the dimensionality imposed by the lattice bonding network and that defined by the band-edge electronic states. Such a structural--electronic dimensionality mismatch provides an intrinsic route to spatially separate the dominant phonon and carrier transport channels without relying on extrinsic defects or structural disorder.

Here, we explore this concept through a unified high-throughput search of Q-1D semiconductors and identify a family of compounds exhibiting strongly suppressed interchain lattice heat transport while retaining favorable interchain hole transport. Among the screened candidates, Sn$_2$S$_3$ and SbTeI are selected as representative systems for detailed microscopic investigation. Sn$_2$S$_3$, an experimentally reported thermoelectric compound independently recovered by our screening, provides an instructive example in which the known anisotropic lattice transport is accompanied by an opposite tendency in hole transport that has not previously been recognized from the perspective of spatial transport decoupling. In both Sn$_2$S$_3$ and SbTeI, strong intrachain bonding favors phonon propagation along the chains, whereas weak interchain interactions strongly suppress transverse lattice heat conduction. In contrast, transversely oriented valence-band $p$ states generate effective electronic coupling between neighboring chains, resulting in relatively light hole transport along the interchain direction. For SbTeI, this inverted anisotropy produces a favorable combination of low interchain lattice thermal conductivity and efficient hole transport, yielding a maximum $zT$ of approximately 2.1 near 900~K. More broadly, the screened candidates exhibit substantially stronger anisotropy in lattice heat transport than in hole transport, with interchain hole mobility frequently comparable to or higher than its intrachain counterpart. These results establish structural--electronic dimensionality mismatch as a general design principle for spatially decoupling phonon and carrier transport in Q-1D thermoelectric materials.

%\tableofcontents

\section*{\Rmnum{2}.COMPUTATIONAL METHODS}

First-principles calculations were performed within density functional theory (DFT) using the Vienna \textit{Ab Initio} Simulation Package (VASP)~\cite{Kresse1996}. The electron--ion interactions were described by the projector augmented-wave (PAW) method, and the exchange--correlation energy was treated using the PBEsol functional~\cite{Perdew2008}. A plane-wave kinetic-energy cutoff of 500~eV was employed. The atomic structures were fully relaxed until the total-energy difference between successive ionic steps was smaller than $10^{-5}$~eV and the residual Hellmann--Feynman forces were below $10^{-3}$~eV\,\AA$^{-1}$. Appropriate $\Gamma$-centered $\mathbf{k}$-point meshes were employed for Brillouin-zone sampling and carefully checked for convergence.

To obtain more reliable electronic band gaps, additional calculations were performed using the screened hybrid HSE06 functional~\cite{Heyd2003}. The HSE06 band gaps were subsequently used to correct the PBEsol band gaps in the electronic-transport calculations, while the underlying band dispersions were obtained from the PBEsol calculations.

Finite-temperature lattice dynamics were investigated using \textit{ab initio} molecular dynamics (AIMD). For Sn$_2$S$_3$ and SbTeI, $2\times5\times1$ and $2\times4\times2$ supercells containing 200 and 192 atoms, respectively, were employed. Owing to the large supercell sizes, the AIMD simulations were sampled at the $\Gamma$ point. The trajectories were propagated with a time step of 1~fs for a total duration of 30~ps. Temperature-dependent interatomic force constants (IFCs) were extracted from the AIMD trajectories using the temperature-dependent effective potential (TDEP) method~\cite{Hellman2013,Knoop2024}. The cutoff radii for the second-, third-, and fourth-order IFCs were set to 10, 6, and 5~\AA, respectively.

The lattice thermal conductivity was calculated by solving the phonon Boltzmann transport equation using ShengBTE~\cite{shengbte2014}. Three-phonon, four-phonon, and isotope scattering processes were included, with the four-phonon scattering rates evaluated using the FourPhonon implementation~\cite{Han2022}. The phonon $\mathbf{q}$-point meshes were carefully tested to ensure convergence. In addition to the conventional population contribution $\kappa_P$ associated with particle-like phonon propagation, the coherence contribution $\kappa_C$ arising from off-diagonal coupling between phonon modes was also evaluated.

Electronic transport properties were calculated using the \textsc{AMSET} package~\cite{Ganose2021}. The HSE06 band gaps were used to correct the PBEsol band gaps in the transport calculations, and an interpolation factor of 12 was employed for electronic-band interpolation. Carrier scattering from ionized impurities (IMP), acoustic deformation potentials (ADP), and polar optical phonons (POP) was included. The electrical conductivity $\sigma$, Seebeck coefficient $S$, electronic thermal conductivity $\kappa_e$, and carrier mobility were evaluated over the relevant ranges of temperature and carrier concentration.

\section*{\Rmnum{3}.RESULTS AND DISCUSSION}

\begin{figure*}[ht!]
\centering
\includegraphics[width=0.8\linewidth]{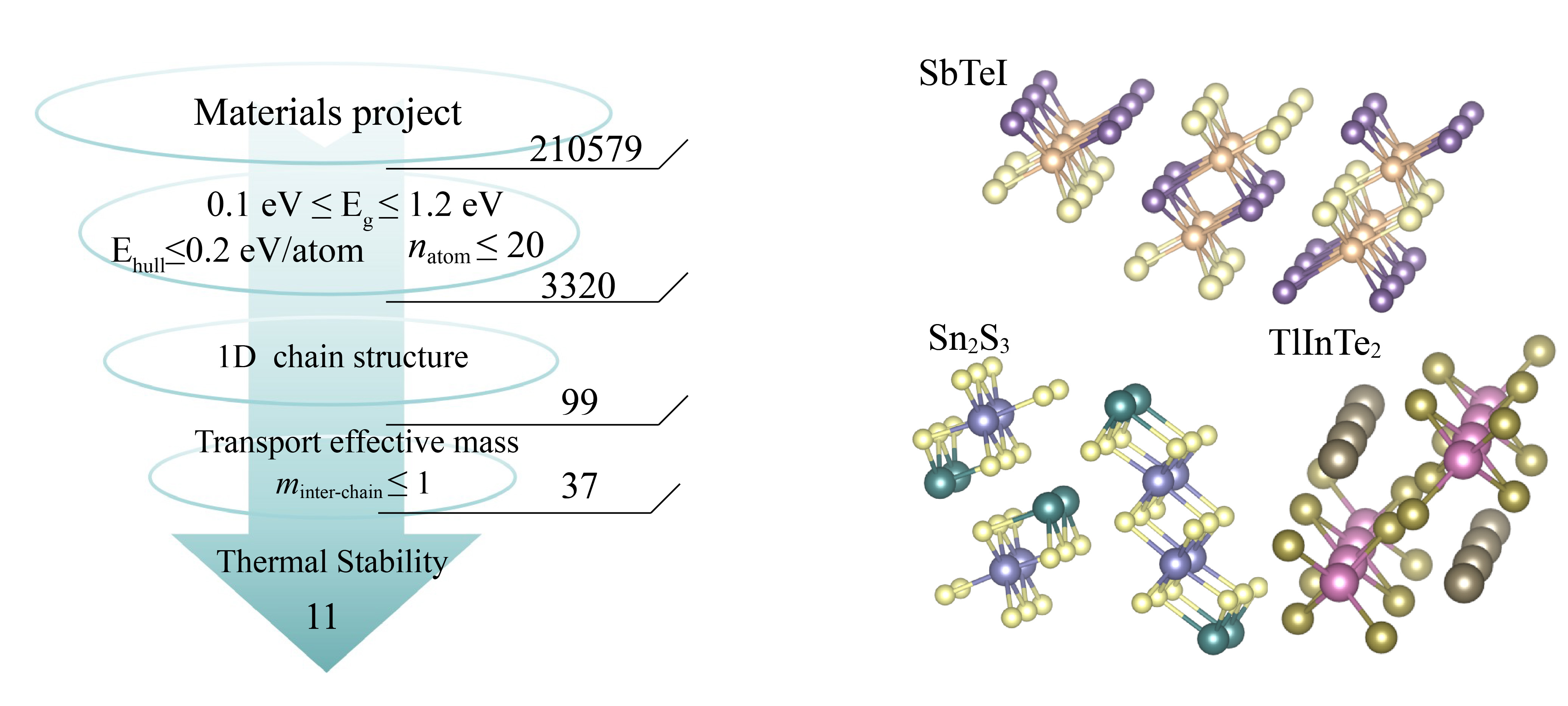}
\caption{High-throughput screening workflow for identifying one-dimensional materials with potential spatially decoupled thermal and electrical transport.}
\label{fig0}
\end{figure*}

We begin with a multistep high-throughput screening of the Materials Project database to identify one-dimensional (1D) materials with the potential for spatially decoupled thermal and electrical transport, as illustrated in Fig.~\ref{fig0}. First, a preliminary screening is performed based on the band gap, thermodynamic stability, and unit-cell size. The remaining structures are then analyzed using periodic bonding networks constructed with CrystalNN, and only materials exhibiting a continuous 1D chain-like connectivity are retained. For these 1D candidates, we further evaluate the direction-dependent conductivity effective masses and select materials with a sufficiently small carrier effective mass along at least one interchain direction. This criterion is designed to identify systems in which weak interchain lattice coupling can coexist with favorable transverse electronic transport. Finally, the remaining candidates are subjected to finite-temperature stability tests at 300~K, yielding a set of promising materials for spatially decoupled thermal and electrical transport.

Among the screened candidates, we first focus on Sn$_2$S$_3$, an experimentally reported compound that is independently recovered by our high-throughput screening. This makes Sn$_2$S$_3$ a particularly instructive prototype for validating the screening strategy. More importantly, although its structural and transport properties have been investigated previously, the contrasting directional preferences of lattice and carrier transport, and the resulting possibility of spatially decoupled thermal and electrical conduction, have not been discussed from this perspective.

\begin{figure*}[ht!]
\centering
\includegraphics[width=1\linewidth]{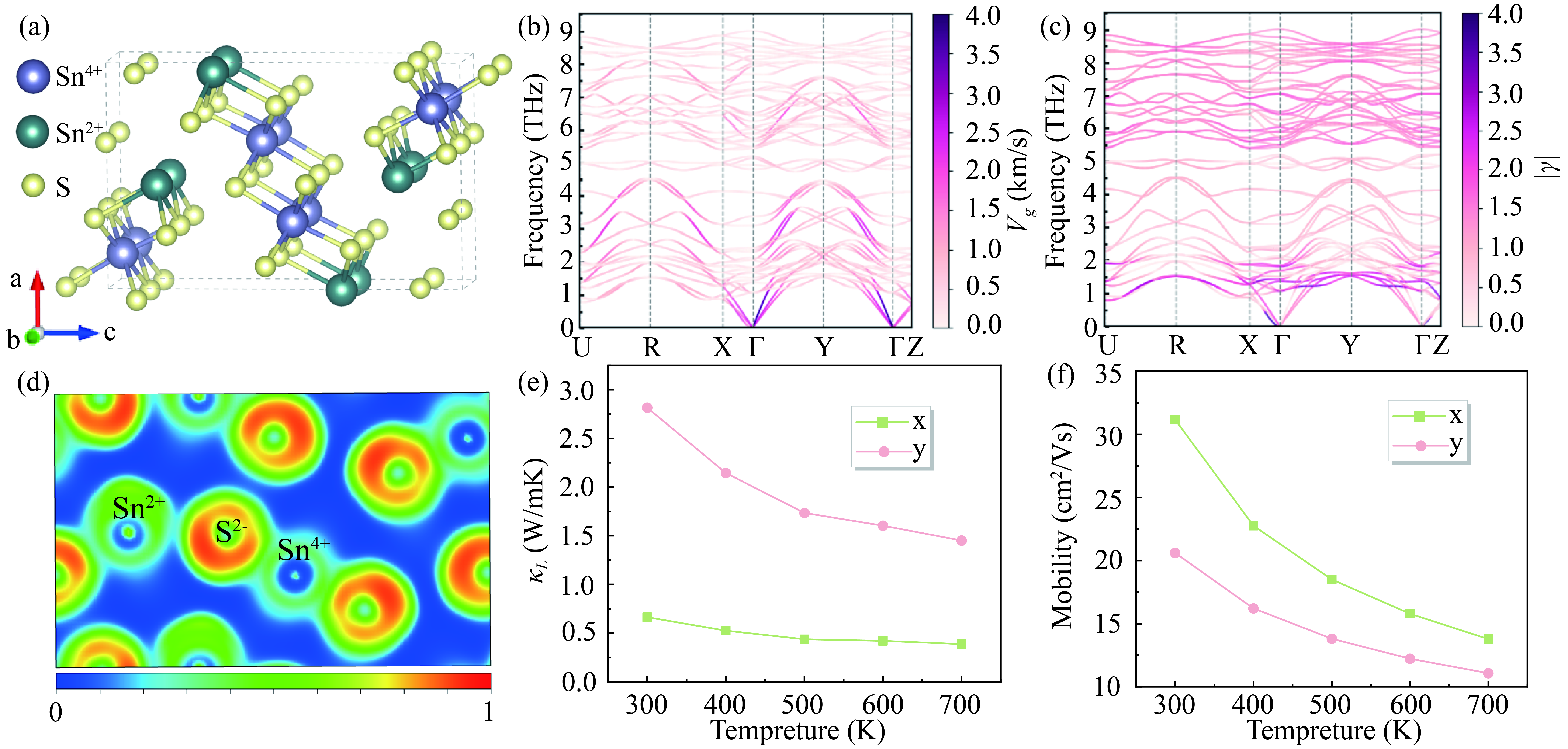}
\caption{
Structural, lattice-dynamical, and transport properties of Sn$_2$S$_3$.
(a) Crystal structure of Sn$_2$S$_3$.
(b) Phonon dispersion colored by the magnitude of the phonon group velocity $|v_g|$.
(c) Phonon dispersion colored by the magnitude of the mode Gr\"uneisen parameter $|\gamma|$.
(d) Electron localization function (ELF).
(e) Temperature-dependent lattice thermal conductivity $\kappa_L$ along the $x$ and $y$ directions including three- and four-phonon scattering.
(f) Temperature-dependent hole mobility along the $x$ and $y$ directions.
}
\label{fig1}
\end{figure*}

As shown in Fig.~\ref{fig1}(a), Sn$_2$S$_3$ crystallizes in the orthorhombic $Pnma$ space group, with 20 atoms in the unit cell. The optimized lattice constants are $a=8.79$~\AA, $b=3.77$~\AA, and $c=13.82$~\AA, in good agreement with experiment\cite{Zhu2025}. Its crystal structure exhibits a pronounced quasi-one-dimensional character\cite{Sun2025,Jiang2025}: the Sn--S framework extends continuously along the $b$ axis, whereas neighboring chains are connected much more weakly along the $a$ and $c$ directions, establishing a clear strong-intrachain--weak-interchain bonding hierarchy. Sn is present in mixed valence states of Sn$^{2+}$ and Sn$^{4+}$.

The resulting lattice-dynamical anisotropy is revealed in Fig.~\ref{fig1}(b) and \ref{fig1}(c). Near the $\Gamma$ point, the average phonon group velocities along $\Gamma$--X, $\Gamma$--Y, and $\Gamma$--Z are approximately 1.69, 2.87, and 1.82~km~s$^{-1}$, respectively. The substantially larger velocity along the intrachain $\Gamma$--Y direction indicates that the strongly bonded Sn--S chains provide an efficient pathway for phonon propagation. Meanwhile, the average magnitude of the mode Gr\"uneisen parameters, $\langle|\gamma|\rangle$, is 1.32, exceeding those of representative thermoelectric materials such as SnS$_2$ ($\sim$1.08), PbS ($\sim$1.18), and SnSe$_2$ ($\sim$1.20), indicative of pronounced intrinsic lattice anharmonicity. In addition, $|\gamma|$ is generally larger along $\Gamma$--X than along $\Gamma$--Y and $\Gamma$--Z, revealing a stronger anharmonic response along the interchain $x$ direction. The coexistence of fast intrachain phonon propagation and enhanced interchain anharmonicity therefore establishes a strongly anisotropic lattice-dynamical landscape.

Further microscopic insight is provided by the electron localization function (ELF) shown in Fig.~\ref{fig1}(d). A spatially directed $5s^2$ cationic lone pair is clearly visible on the Sn$^{2+}$ site, producing an asymmetric and relatively soft local bonding environment. Such local electronic asymmetry weakens the restoring force against atomic displacement and facilitates large-amplitude, low-frequency rattling-like vibrations, thereby enhancing lattice anharmonicity and phonon scattering.

These lattice-dynamical characteristics translate directly into strongly anisotropic heat transport. As shown in Fig.~\ref{fig1}(e), when both three- and four-phonon scattering processes are included, the lattice thermal conductivity at 300~K reaches 2.82~W~m$^{-1}$~K$^{-1}$ along the intrachain $y$ direction, but is only 0.66~W~m$^{-1}$~K$^{-1}$ along the interchain $x$ direction. The former is therefore about 4.27 times larger than the latter, demonstrating that heat is preferentially transported along the strongly bonded chains.

Remarkably, the carrier transport exhibits the opposite directional preference. As shown in Fig.~\ref{fig1}(f), the hole mobility at 300~K is 31.17~cm$^2$~V$^{-1}$~s$^{-1}$ along the interchain $x$ direction, compared with 20.61~cm$^2$~V$^{-1}$~s$^{-1}$ along the intrachain $y$ direction. Sn$_2$S$_3$ therefore exhibits an inverted thermal--electrical transport anisotropy: phonons preferentially propagate along the chains, whereas holes are transported more efficiently across them. In particular, the interchain $x$ direction simultaneously combines a low lattice thermal conductivity with a relatively high hole mobility, providing a direct manifestation of spatially decoupled phonon and carrier transport.

\begin{figure*}[ht!]
\centering
\includegraphics[width=0.8\linewidth]{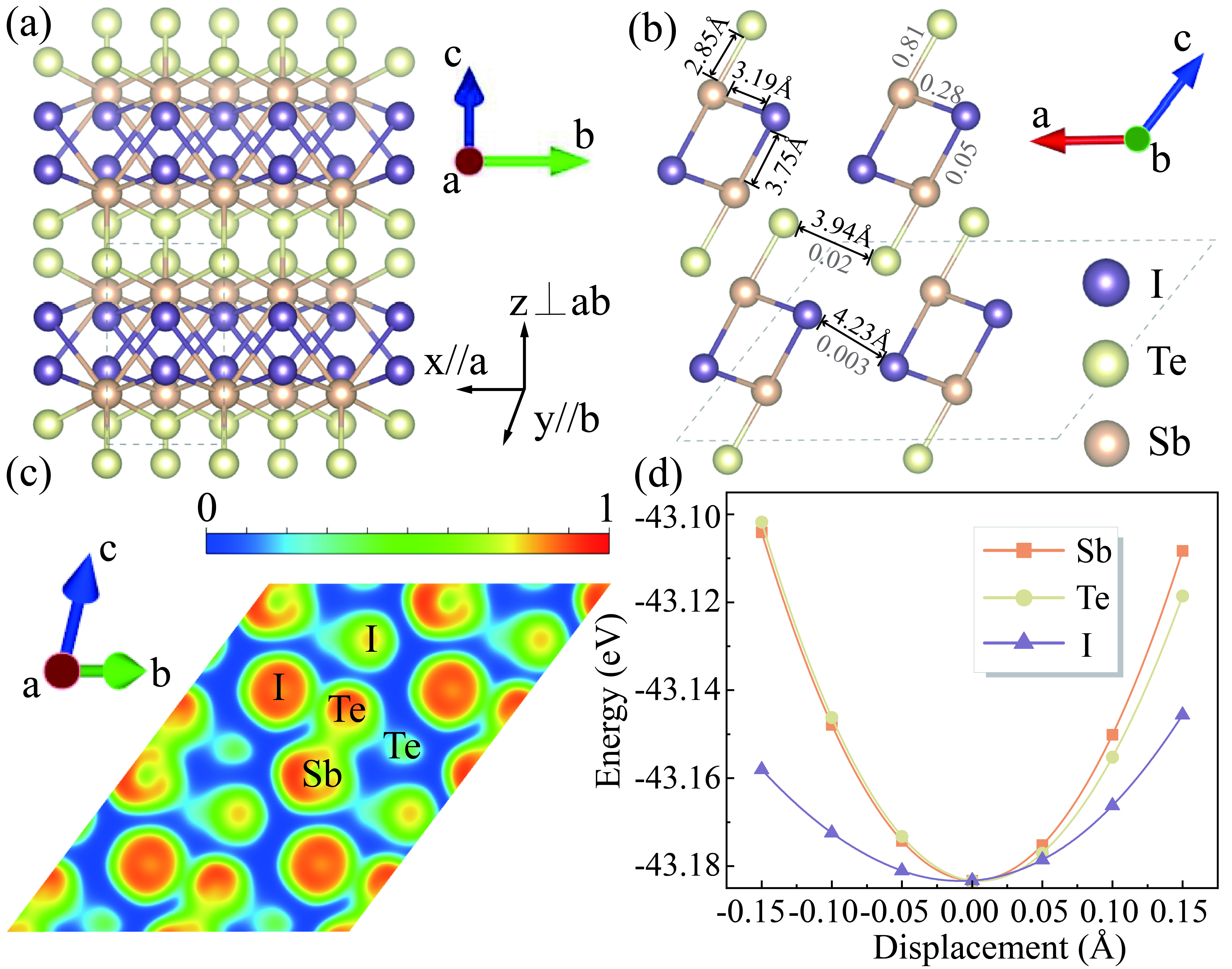}
\caption{
Crystal structure and local bonding characteristics of SbTeI.
(a) Crystal structure of SbTeI.
(b) Structure projected along the $b$ axis onto the $ac$ plane, with representative interatomic distances and ICOBI values.
(c) Electron localization function (ELF).
(d) Potential-energy profiles for displacements of Sb, Te, and I atoms along the $a$ direction.
}
\label{fig2}
\end{figure*}

Having established the inverted thermal--electrical transport anisotropy in Sn$_2$S$_3$, we next turn to SbTeI, another candidate identified by our screening, to examine how the same strong-intrachain--weak-interchain structural hierarchy governs its lattice response. SbTeI crystallizes in a base-centered monoclinic structure with the space group $C2/m$ (No.~12). The optimized lattice constants are $a=13.77$~\AA, $b=4.21$~\AA, and $c=9.16$~\AA, with $\alpha=\gamma=90^\circ$ and $\beta=128.97^\circ$, in good agreement with experiment\cite{Sereika2022}. As shown in Fig.~\ref{fig2}(a), its fundamental structural motif consists of one-dimensional double chains extending continuously along the short $b$ axis, with the intrachain framework primarily connected by Sb--Te and Sb--I bonds. Neighboring double chains are further arranged side by side into plate-like blocks, whereas the interactions between these blocks are considerably weaker, giving rise to a pronounced intrachain--interchain bonding hierarchy.

Figure~\ref{fig2}(b) further shows the structure projected along the $b$ axis onto the $ac$ plane, together with the bonding characteristics of representative atomic pairs. Within the chains, the nearest-neighbor Sb--Te bond has a length of 2.85~\AA\ and an ICOBI value of 0.81, indicating pronounced covalent bonding, whereas the Sb--I bond is longer, 3.19~\AA, with a smaller ICOBI of 0.28. In sharp contrast, the interchain Te--Te and I--I separations reach 3.94 and 4.23~\AA, respectively, with corresponding ICOBI values of only 0.02 and 0.003, indicating negligible direct bonding between neighboring chains. These results quantitatively establish the strong-intrachain--weak-interchain bonding hierarchy in SbTeI and provide a microscopic structural basis for its anisotropic lattice response.

The electron localization function shown in Fig.~\ref{fig2}(c) provides further insight into this anisotropic bonding environment. A spatially directed $5s^2$ cationic lone pair is clearly associated with Sb$^{3+}$, with its electron density extending predominantly toward the open space between neighboring chains. Experimental structural studies have likewise suggested that Sb in SbTeI resides in an incomplete local coordination environment, with the lone-pair electrons preferentially occupying the interstitial region between adjacent plate-like blocks. Such asymmetric electron localization further enhances the directional character and local softness of the bonding environment around Sb, thereby favoring anharmonic lattice vibrations.

To quantify the local lattice softening associated with the weak interchain bonding, Fig.~\ref{fig2}(d) shows the potential-energy variations upon displacing Sb, Te, and I atoms along the $a$ direction from their equilibrium positions. Among the three species, I exhibits the flattest potential-energy profile and the lowest energy cost for displacement, indicating the smallest effective restoring force and the weakest local lattice constraint. This behavior is fully consistent with the weak interchain bonding inferred from the bond lengths and ICOBI analysis. It demonstrates that the interchain direction in SbTeI is characterized by a particularly soft local potential-energy landscape, providing a microscopic origin for the suppressed transverse phonon transport and the strong thermal-transport anisotropy discussed below.

\begin{figure*}[ht!]
\centering
\includegraphics[width=1\linewidth]{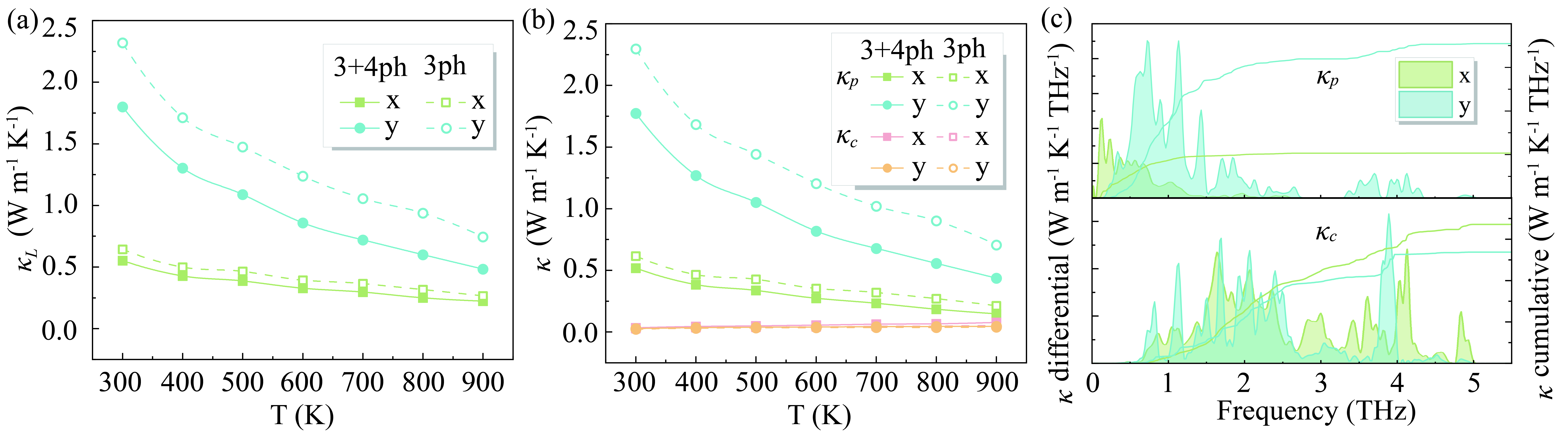}
\caption{
Lattice thermal transport properties of SbTeI.
(a) Temperature-dependent lattice thermal conductivity $\kappa_L$ along the $x$ and $y$ directions calculated with three-phonon (3ph) and combined three- and four-phonon (3+4ph) scattering.
(b) Population ($\kappa_P$) and coherence ($\kappa_C$) contributions to the lattice thermal conductivity along the $x$ and $y$ directions.
(c) Frequency-resolved differential and cumulative contributions to $\kappa_P$ and $\kappa_C$ along the $x$ and $y$ directions.
}
\label{fig3}
\end{figure*}

Consistent with the soft interchain potential-energy landscape discussed above, SbTeI exhibits strongly anisotropic lattice heat transport. As shown in Fig.~\ref{fig3}(a), the lattice thermal conductivities along both the $x$ and $y$ directions decrease with increasing temperature, with $\kappa_L$ along the interchain $x$ direction remaining substantially lower than that along the intrachain $y$ direction over the entire temperature range. This anisotropy directly reflects the strong-intrachain--weak-interchain bonding hierarchy: the continuous Sb--Te and Sb--I bonding network along $y$ facilitates phonon propagation, whereas the much weaker interchain interactions suppress heat transport along $x$. At 300~K, inclusion of four-phonon scattering reduces $\kappa_{L,x}$ from 0.64 to 0.55~W\,m$^{-1}$\,K$^{-1}$ and $\kappa_{L,y}$ from 2.32 to 1.80~W\,m$^{-1}$\,K$^{-1}$, corresponding to reductions of approximately 14\% and 22\%, respectively, and yielding an anisotropy ratio of $\kappa_{L,y}/\kappa_{L,x}=3.27$. The increasing importance of this higher-order scattering at elevated temperatures highlights the pronounced anharmonicity of SbTeI.

To further elucidate the heat-transport mechanism, Fig.~\ref{fig3}(b) decomposes $\kappa_L$ into the population contribution $\kappa_P$, associated with particle-like phonon propagation, and the coherence contribution $\kappa_C$, arising from off-diagonal coupling between phonon modes. Heat conduction is dominated by $\kappa_P$, which decreases markedly with increasing temperature. In contrast, along the interchain $x$ direction, $\kappa_C$ changes only from approximately 0.03~W\,m$^{-1}$\,K$^{-1}$ at 300~K to 0.08~W\,m$^{-1}$\,K$^{-1}$ at 900~K. As a result, its relative contribution progressively increases and reaches approximately 34\% at 900~K.

This growing relative importance of the coherence channel weakens the temperature dependence of the total lattice thermal conductivity. In particular, the interchain component follows approximately
$\kappa_{L,x}\propto T^{-0.77}$,
deviating from the conventional $T^{-1}$ dependence commonly associated with Umklapp-dominated three-phonon transport. A similar behavior has been reported in AgSbTe$_2$\cite{Kwon2026}, where a comparatively weakly temperature-dependent off-diagonal contribution gains relative weight as the particle-like contribution is suppressed.

The frequency-resolved thermal conductivity in Fig.~\ref{fig3}(c) further distinguishes these two transport channels. The population contribution is concentrated primarily in the low-frequency region, indicating that low-frequency phonons carry most of the particle-like heat current. By contrast, the coherence contribution extends over a broader frequency range, consistent with coupling between phonon modes of similar frequencies. The broad spectral distribution of $\kappa_C$ therefore provides a microscopic basis for its increasing relative importance at elevated temperatures and the weakened temperature dependence of the interchain thermal conductivity.

\begin{figure*}[ht!]
\centering
\includegraphics[width=1\linewidth]{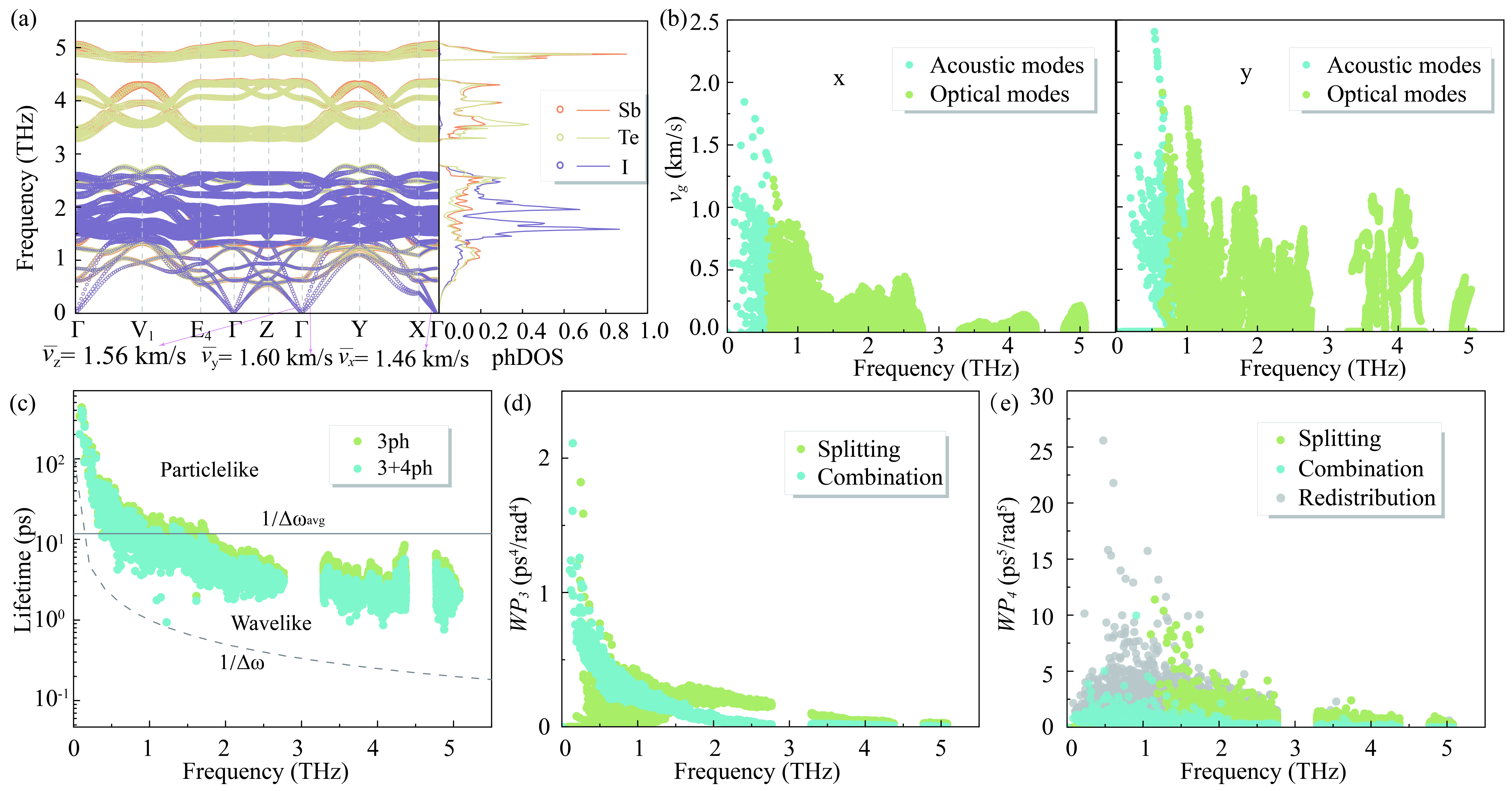}
\caption{
Lattice-dynamical and anharmonic scattering characteristics of SbTeI.
(a) Atom-projected phonon dispersion and phonon density of states (phDOS).
(b) Frequency-dependent phonon group velocities along the $x$ and $y$ directions, resolved into acoustic and optical modes.
(c) Phonon lifetimes calculated with three-phonon (3ph) and combined three- and four-phonon (3+4ph) scattering; the Ioffe--Regel and Wigner characteristic times are also indicated.
(d) Three-phonon scattering phase space $W_3$, resolved into splitting and combination processes.
(e) Four-phonon scattering phase space $W_4$, resolved into splitting, combination, and redistribution processes.
}
\label{fig4}
\end{figure*}

To further uncover the microscopic origin of the anisotropic thermal transport in SbTeI, Fig.~\ref{fig4}(a) presents the atomic projections of the phonon eigenvectors onto the phonon spectrum, where $\Gamma$--X, $\Gamma$--Y, and $\Gamma$--Z correspond to the Cartesian $x$, $y$, and $z$ directions, respectively. Combined with the phonon density of states, the results show that the low-frequency vibrations are dominated by I atoms, whereas Sb and Te contribute primarily at higher frequencies. In particular, the I-derived vibrations are concentrated within a relatively narrow low-frequency window, reflecting the weak local confinement of I atoms and consistent with the weak bonding and soft potential-energy landscape discussed above. The average phonon group velocities near $\Gamma$ along $\Gamma$--X, $\Gamma$--Y, and $\Gamma$--Z are approximately 1.46, 1.60, and 1.56~km~s$^{-1}$, respectively, with the largest value along the intrachain $\Gamma$--Y direction. As further shown in Fig.~\ref{fig4}(b), the phonon group velocities along $y$ are generally larger than those along $x$, providing a dynamical manifestation of the strong-intrachain--weak-interchain bonding hierarchy: strong intrachain bonding promotes phonon propagation, whereas weak interchain coupling limits transverse heat transport.

The dense phonon spectrum also provides favorable conditions for the coherence contribution discussed above. Comparing Fig.~\ref{fig4}(a) with the frequency-resolved thermal conductivity in Fig.~\ref{fig3}(c), the frequency regions with a high phonon density of states broadly coincide with those in which $\kappa_C$ is distributed. A dense vibrational spectrum increases the number of phonon branches with small frequency separations, thereby providing more nearly degenerate mode pairs that can contribute through off-diagonal intermode coupling. This explains why, in contrast to the population contribution $\kappa_P$, which is concentrated mainly at low frequencies, the coherence contribution $\kappa_C$ extends over a much broader frequency range.

Further evidence for the coexistence of particle-like and wave-like transport is provided by the phonon lifetimes in Fig.~\ref{fig4}(c). The two reference curves denote the Ioffe--Regel time, $\tau_{\mathrm{IR}}=\omega^{-1}$, and the Wigner time, $\tau_{\mathrm{W}}=(\Delta\omega_{\mathrm{avg}})^{-1}$, where $\Delta\omega_{\mathrm{avg}}$ represents the average frequency spacing between neighboring phonon branches. A substantial fraction of the phonon modes falls in the intermediate regime between these two characteristic times, where the conventional particle-like phonon picture becomes progressively less dominant and interband coherence becomes increasingly relevant. The dense and closely spaced phonon branches in SbTeI therefore provide a microscopic basis for the finite coherence contribution to the lattice thermal conductivity identified in Fig.~\ref{fig3}.

The strong anharmonic scattering is further reflected in the three- and four-phonon scattering phase spaces, $W_3$ and $W_4$, shown in Fig.~\ref{fig4}(d) and \ref{fig4}(f), respectively. For three-phonon processes, combination channels dominate below approximately 1.5~THz, whereas the available phase space for splitting processes increases markedly at higher frequencies. For four-phonon scattering, the redistribution channel exhibits a pronounced enhancement near 1~THz, coinciding with the low-frequency, relatively flat phonon branches dominated by I vibrations. These flat and densely packed phonon branches provide abundant energy- and momentum-conserving channels for multiphonon scattering, thereby enhancing anharmonic phonon scattering and further suppressing the lattice thermal conductivity of SbTeI.

The results above establish a common lattice-transport picture for Sn$_2$S$_3$ and SbTeI: their thermal transport is strongly governed by the strong-intrachain--weak-interchain bonding hierarchy, such that phonons propagate preferentially along the strongly bonded chains while heat conduction is strongly suppressed across the chains. This behavior contrasts with the carrier-transport anisotropy already observed in Sn$_2$S$_3$, where the hole mobility is higher along the interchain direction. To uncover the electronic origin of this inverted transport anisotropy, we next examine the band structures, orbital characters near the valence-band maximum (VBM), effective masses, chemical bonding, and real-space VBM charge distributions of both compounds.

\begin{figure*}[ht!]
\centering
\includegraphics[width=1\linewidth]{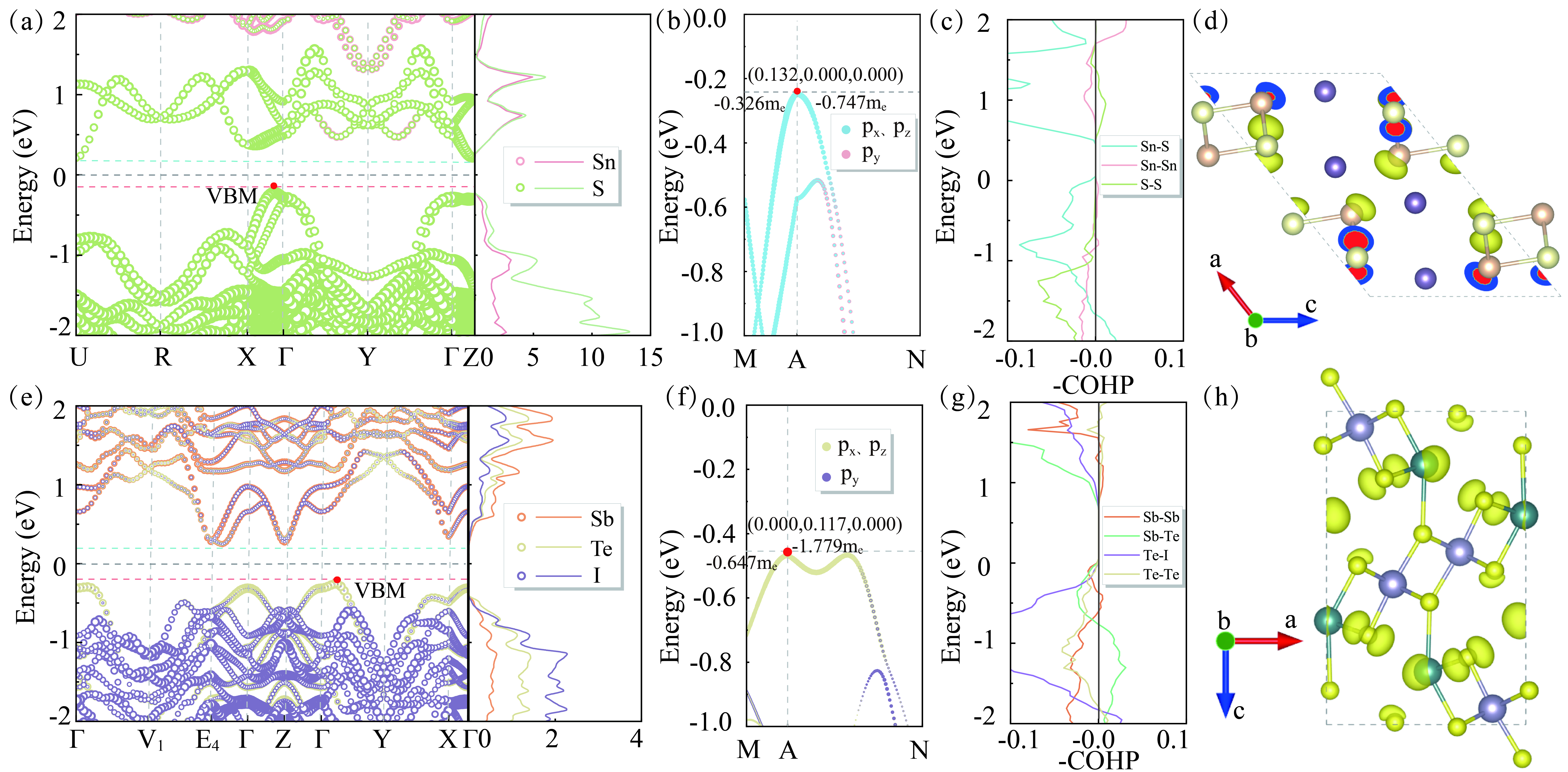}
\caption{Electronic structures and valence-band characteristics of Sn$_2$S$_3$ and SbTeI.
(a,e) Orbital-resolved band structures and corresponding density of states of Sn$_2$S$_3$ and SbTeI, respectively.
(b,f) Local M--A--N band dispersions around the valence-band maximum (VBM), where M--A and A--N correspond to the interchain $x$ and intrachain $y$ directions, respectively.
(c,g) $-\mathrm{COHP}$ curves for representative bonding interactions.
(d,h) Real-space charge densities of the VBM states.}
\label{fig5}
\end{figure*}

Within PBE, Sn$_2$S$_3$ and SbTeI are both indirect-gap semiconductors, with band gaps of 0.421 and 0.507~eV, respectively. As shown in Fig.~\ref{fig5}(a) and \ref{fig5}(e), the VBM of Sn$_2$S$_3$ lies along the $\Gamma$--X path, whereas that of SbTeI occurs along $\Gamma$--Y. Because hole transport is governed primarily by the electronic states close to the VBM, we focus on the local band dispersion and orbital character in the vicinity of these extrema.

To resolve the directional dispersion around the actual VBM, we construct local $\mathbf{k}$ paths passing through each VBM and label the band-edge point as A. The resulting M--A--N paths, shown in Fig.~\ref{fig5}(b) and \ref{fig5}(f), are chosen such that M--A and A--N probe the interchain $x$ and intrachain $y$ directions, respectively. The orbital-resolved band structures in Fig.~\ref{fig5}(a) and \ref{fig5}(e) show that the VBM of Sn$_2$S$_3$ is dominated by S-$p_x/p_z$ states, whereas that of SbTeI is primarily derived from Te-$p_x/p_z$ states. Since the one-dimensional chains in both materials extend along the $y$ direction, these orbital components lie predominantly in the plane transverse to the chain axis and provide electronic weight in the interchain region. They can therefore support substantial orbital overlap and electronic coupling between neighboring chains. This is a key distinction between the structural and electronic dimensionalities of these compounds: although the crystal frameworks are strongly quasi-one-dimensional, the frontier valence states are not confined to the chain direction but retain pronounced transverse character.

This orbital anisotropy is directly reflected in the VBM dispersion and the direction-dependent hole effective masses [Fig.~\ref{fig5}(b) and \ref{fig5}(f)]. For Sn$_2$S$_3$, the hole band effective masses along the interchain $x$ and intrachain $y$ directions are $0.326m_0$ and $0.747m_0$, respectively, indicating a substantially larger band curvature along $x$. At 300~K and a hole concentration of $10^{18}$~cm$^{-3}$, the corresponding conductivity effective masses are $0.58m_0$ and $0.92m_0$. Both measures therefore consistently identify the interchain $x$ direction as the electronically lighter transport direction, in agreement with the higher hole mobility found along $x$.

SbTeI exhibits the same qualitative trend. Its hole band effective masses are $0.657m_0$ along $x$ and $1.78m_0$ along $y$, demonstrating a much stronger valence-band dispersion across the chains than along them. The conductivity effective masses at 300~K and $10^{18}$~cm$^{-3}$ are correspondingly $0.94m_0$ and $1.02m_0$ along $x$ and $y$, respectively. Although the anisotropy is less pronounced in the transport effective mass than in the local band curvature, both quantities preserve the relation
\[
m_x^* < m_y^*,
\]
showing that the interchain direction remains favorable for hole transport in both compounds.

Additional information on the bonding character of the VBM states is provided by the $-\mathrm{COHP}$ analysis in Fig.~\ref{fig5}(c) and \ref{fig5}(g). In both Sn$_2$S$_3$ and SbTeI, the states near the VBM contain appreciable antibonding character, indicating that the frontier valence states are not simply localized on the strongest intrachain bonds. Such antibonding character makes these states particularly sensitive to local bond distortions and is consistent with the relatively soft and anharmonic bonding environments identified from the lattice-dynamical analysis. Importantly, this bonding character coexists with relatively light hole masses, showing that electronic delocalization across the chains can be maintained even in a lattice with weak transverse mechanical coupling.

The real-space VBM charge densities shown in Fig.~\ref{fig5}(d) and \ref{fig5}(h) provide a more direct visualization of this behavior. In both compounds, the VBM charge exhibits pronounced directional extension into the interchain region, with appreciable overlap between orbitals on neighboring chains, whereas the electronic distribution along the intrachain $y$ direction is comparatively more localized. This spatial distribution is consistent with the dominant S-$p_x/p_z$ and Te-$p_x/p_z$ orbital characters and with the smaller hole effective masses along $x$. The transverse $p$ orbitals therefore establish effective electronic coupling pathways across neighboring chains, despite the much weaker interchain lattice bonding.

These results reveal that the anisotropies of phonon and carrier transport in Sn$_2$S$_3$ and SbTeI originate from fundamentally different microscopic factors. Phonon transport is governed primarily by the mechanical bonding hierarchy: strong intrachain bonds produce a stiffer lattice and higher phonon group velocities, favoring heat flow along the chains, whereas weak interchain interactions and enhanced anharmonic scattering suppress transverse lattice heat transport. Hole transport, by contrast, is controlled by the orbital character and dispersion of the VBM states. The transverse extension and interchain hybridization of S-$p_x/p_z$ and Te-$p_x/p_z$ states generate smaller hole effective masses and more favorable carrier transport along the interchain $x$ direction. Consequently, both materials exhibit an inverted thermal--electrical transport anisotropy, with heat preferentially conducted along the chains while holes can be transported more efficiently across them. This separation of the preferred phonon and carrier transport directions constitutes the microscopic origin of their spatially decoupled thermal and electrical transport.

\begin{figure*}[ht!]
\centering
\includegraphics[width=1\linewidth]{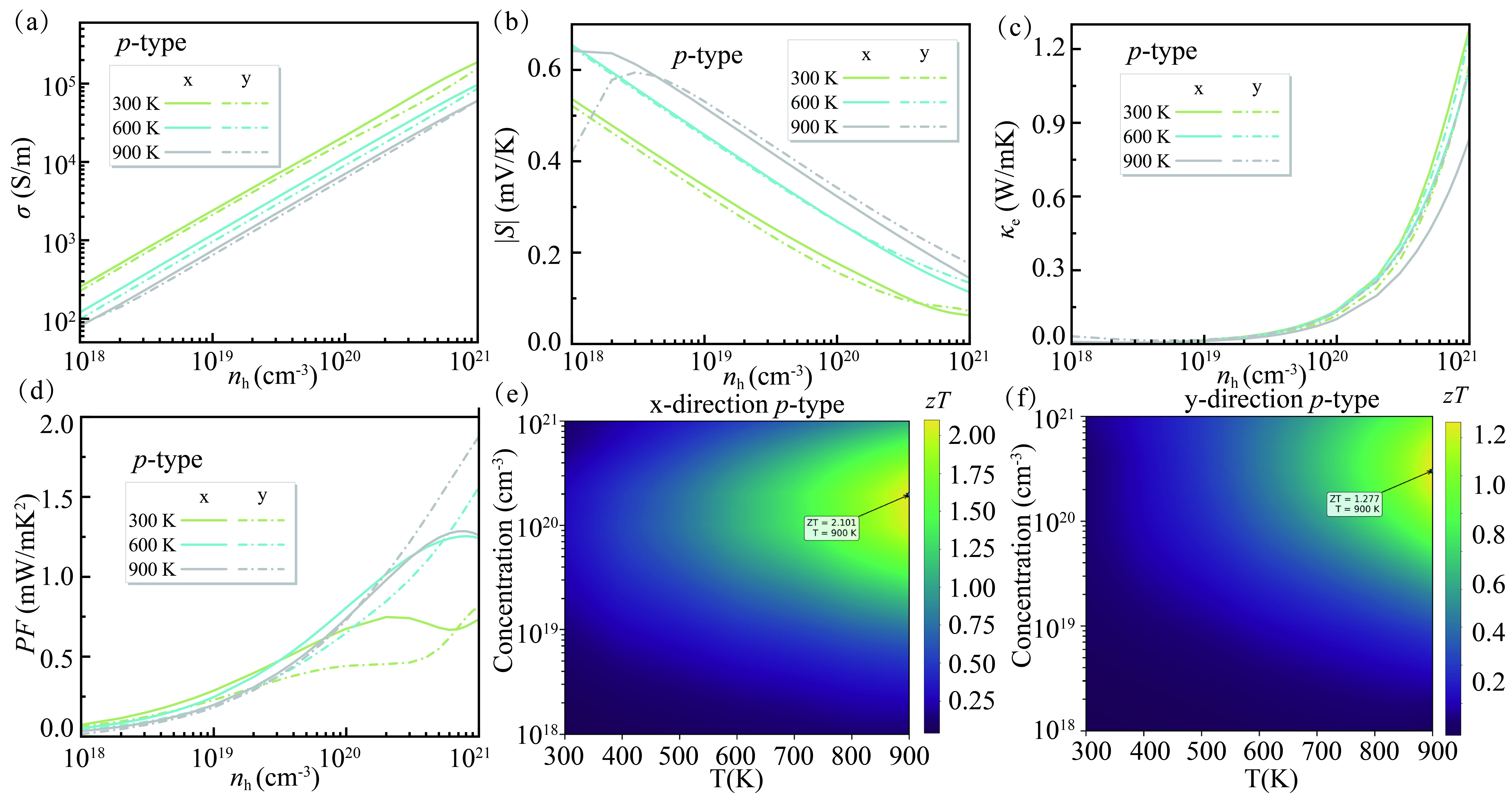}
\caption{
$p$-type thermoelectric transport properties of SbTeI as functions of hole concentration and temperature.
(a) Electrical conductivity $\sigma$.
(b) Seebeck coefficient $S$.
(c) Electronic thermal conductivity $\kappa_e$.
(d) Power factor $PF=S^2\sigma$.
(e,f) Thermoelectric figure of merit $zT$ along the $x$ and $y$ directions, respectively.
}
\label{fig6}
\end{figure*}

The spatial decoupling between phonon and carrier transport discussed above has direct consequences for the thermoelectric performance of SbTeI. Figure~\ref{fig6} summarizes its $p$-type transport properties over a range of temperatures and hole concentrations. As shown in Fig.~\ref{fig6}(a), the electrical conductivity increases monotonically with hole concentration along both the $x$ and $y$ directions, while exhibiting only weak directional anisotropy. For example, at 300~K and $n_h=10^{19}$~cm$^{-3}$, $\sigma_x$ and $\sigma_y$ are 2396.39 and 2122.45~S\,m$^{-1}$, respectively, corresponding to an anisotropy ratio of only $\sigma_x/\sigma_y=1.13$. Thus, in contrast to the pronounced anisotropy of the lattice thermal conductivity, the electrical conductivity remains relatively comparable along the two directions.

The Seebeck coefficient, shown in Fig.~\ref{fig6}(b), also displays only modest anisotropy. At 300~K, $S_x$ is slightly larger than $S_y$, whereas the relation reverses at elevated temperatures. For instance, at $n_h=10^{20}$~cm$^{-3}$, $S_x$ and $S_y$ are approximately 0.18 and 0.16~mV\,K$^{-1}$ at 300~K, and increase to 0.32 and 0.34~mV\,K$^{-1}$ at 900~K, respectively. The electronic thermal conductivity $\kappa_e$ [Fig.~\ref{fig6}(c)] increases with carrier concentration in both directions, following the enhanced electronic heat transport associated with increasing electrical conductivity. The combined evolution of $S$ and $\sigma$ gives rise to the power factor $PF=S^2\sigma$, shown in Fig.~\ref{fig6}(d). At 900~K and $n_h=10^{21}$~cm$^{-3}$, the power factor reaches 1.88~mW\,m$^{-1}$\,K$^{-2}$ along $y$, compared with 1.26~mW\,m$^{-1}$\,K$^{-2}$ along $x$.

Importantly, however, the thermoelectric figure of merit is determined by the balance between electronic transport and the total thermal conductivity,
\[
zT=\frac{S^2\sigma T}{\kappa_e+\kappa_L}.
\]
Although the $y$ direction exhibits a somewhat larger power factor at high carrier concentrations, this advantage is offset by its substantially higher lattice thermal conductivity. In contrast, the interchain $x$ direction combines a strongly suppressed $\kappa_L$ with only a modest reduction in electronic transport performance. As shown in Fig.~\ref{fig6}(e) and \ref{fig6}(f), this favorable balance leads to markedly enhanced thermoelectric performance along $x$, with the maximum $zT$ reaching approximately 2.1 near 900~K and exceeding that along the intrachain $y$ direction. These results demonstrate that the inverted transport anisotropy of SbTeI---efficient heat conduction along the chains but comparatively favorable carrier transport across them---provides an effective route to decouple phonon and electronic transport and thereby optimize thermoelectric performance.

\begin{figure*}[ht!]
\centering
\includegraphics[width=1\linewidth]{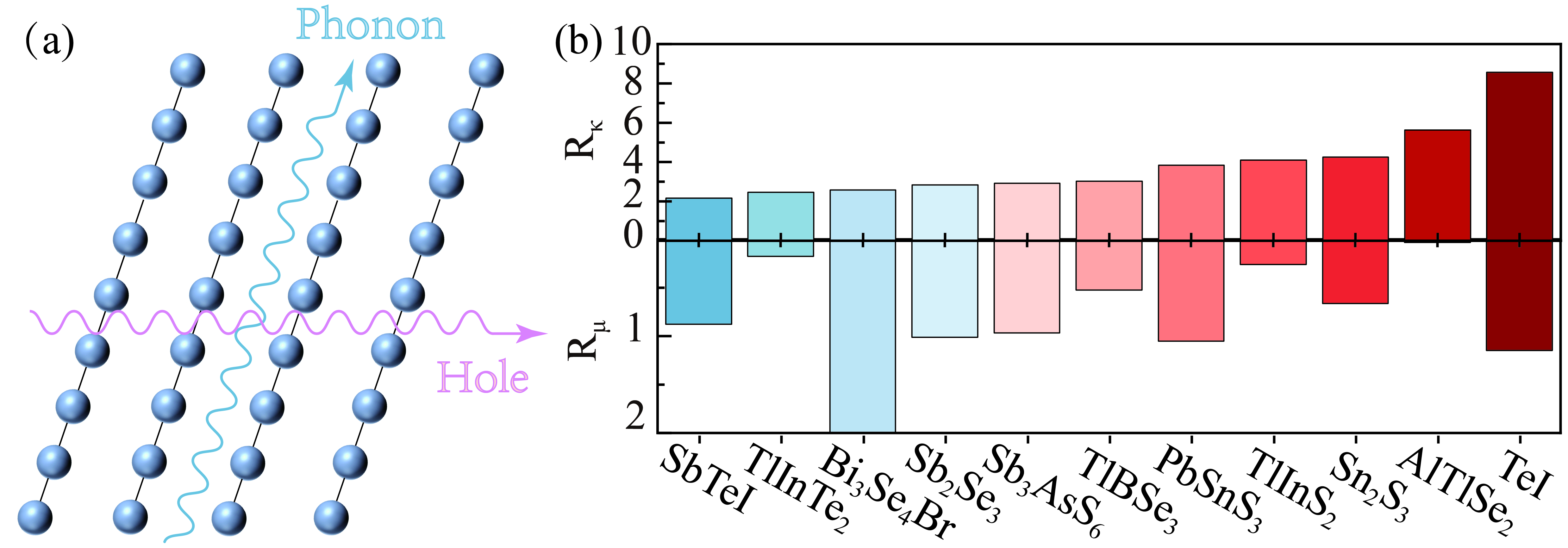}
\caption{
Thermal--electrical transport decoupling in the screened one-dimensional materials.
(a) Schematic illustration of preferential intrachain phonon transport and interchain hole transport in a quasi-one-dimensional structure.
(b) Thermal- and hole-transport anisotropy ratios, $R_{\kappa}=\kappa_L^{\mathrm{inter}}/\kappa_L^{\mathrm{intra}}$ and $R_{\mu}=\mu_h^{\mathrm{inter}}/\mu_h^{\mathrm{intra}}$, for the 11 screened candidates.
}
\label{fig7}
\end{figure*}

To assess whether this transport decoupling extends beyond the representative compounds discussed above, we further compare the directional thermal and hole transport of the 11 candidates identified through the high-throughput screening. The underlying transport picture is illustrated schematically in Fig.~\ref{fig7}(a), where phonons preferentially propagate along the strongly bonded chains, whereas holes can retain efficient transport across the weakly coupled interchain direction.

To quantify this contrasting anisotropy, we define
\[
R_{\kappa}=\frac{\kappa_L^{\mathrm{inter}}}
{\kappa_L^{\mathrm{intra}}},
\qquad
R_{\mu}=\frac{\mu_h^{\mathrm{inter}}}
{\mu_h^{\mathrm{intra}}}.
\]
As summarized in Fig.~\ref{fig7}(b), $R_{\kappa}$ ranges approximately from 0.125 to 0.5 for the screened materials, reflecting a pronounced suppression of lattice heat transport across the chains. In contrast, $R_{\mu}$ exhibits a much weaker anisotropy and exceeds unity for most candidates, indicating that the interchain hole mobility is comparable to, or even higher than, its intrachain counterpart. The markedly different behaviors of $R_{\kappa}$ and $R_{\mu}$ demonstrate that strong structural one-dimensionality does not impose a corresponding one-dimensionality on carrier transport. Instead, suppressed interchain phonon transport can coexist with efficient transverse carrier conduction, providing a general route toward spatially decoupled thermal and electrical transport.

\section*{\Rmnum{4}.CONCLUSIONS}

In summary, we establish a microscopic picture for spatially decoupled thermal and electrical transport in quasi-one-dimensional materials. High-throughput screening identifies a class of compounds in which strong intrachain bonding promotes phonon propagation along the chains, whereas weak interchain coupling strongly suppresses transverse lattice heat transport. Sn$_2$S$_3$ and SbTeI further reveal that carrier transport does not necessarily follow this structural one-dimensionality. Their valence-band-edge states possess pronounced transverse $p$-orbital character, enabling effective interchain electronic coupling and relatively light hole transport across the chains. As a result, the preferred directions of phonon and carrier transport become partially or even fully inverted.

This behavior extends beyond the two representative compounds. Across the screened candidates, the interchain lattice thermal conductivity is consistently suppressed relative to the intrachain direction, while the hole mobility remains much less anisotropic and frequently favors interchain transport. The enhanced thermoelectric performance of SbTeI, with a maximum $zT$ of approximately 2.1 near 900~K, further demonstrates the benefit of such transport decoupling. These results show that structural one-dimensionality need not imply electronic one-dimensionality and establish the mismatch between lattice and electronic transport dimensionalities as a useful design principle for thermoelectric materials.

\section*{Acknowledgements}
This work is supported by the Natural Science Foundation of China (Grants No. 12304038, 52206092, 12204402), the National Key Research and Development Program of China (Grants No. 2024YFA1409800), the Big Data Computing Center of Southeast University and the Center for Fundamental and Interdisciplinary Sciences of Southeast University, Basic Research Program of Jiangsu (BK20250035), Major Basic Research Project of the Natural Science Foundation of the Jiangsu Higher Education Institutions (25KJA470006).


%\bibliography{new1}% Produces the bibliography via BibTeX.

\end{document}